\documentclass[twocolumn,showpacs,preprintnumbers,amsmath,amssymb]{revtex4}
\usepackage{graphicx}% Include figure files
\usepackage{dcolumn}% Align table columns on decimal point
\usepackage{bm}% bold math

\begin{document}

\preprint{}

\title{Effect of Local Channels on Quantum Steering Ellipsoids}% Force line breaks with \\

\author{Xueyuan Hu}
\email{xyhu@sdu.edu.cn}
\affiliation{School of Information Science and Engineering, Shandong University, Jinan 250100, China}%Lines break automatically or can be forced with \\
\author{Heng Fan}
%\author{D. L. Zhou}
%\author{Wu-Ming Liu}
\affiliation{Beijing National Laboratory for Condensed Matter
Physics, Institute of Physics, Chinese Academy of Sciences, Beijing
100190, China}
\date{\today}% It is always \today, today,

\begin{abstract}
The effect of a local trace-preserving single-qubit channel on a two-qubit state is investigated in the picture of the quantum steering ellipsoids (QSE). The phenomenon of locally increased quantum correlation is visualized in this picture. We strictly prove that a $B$-side two-qubit discordant state can be locally prepared from a classical state by a trace-preserving channel on qubit $B$ if and only if its QSE of $B$ is a non-radial line segment. For states with higer-dimensional QSEs, the phenomenon of locally increased quantum correlation generally happens when the shape of the QSE is like a baguette. Based on this observation, we find a class of entangled states whose quantum discord can be increased by the local amplitude damping channel. Further, We find that the local quantum channel does not increase the size of QSEs of either qubit $A$ or qubit $B$, for the needle-shape QSE states, as well as the Bell diagonal states with higher-dimensional QSE.
\end{abstract}

\maketitle

\section{Introduction}
The quantum steering ellipsoid of a two-qubit state is the whole set of Bloch vectors that the qubit $A$ can be collapsed to by a positive-operator valued measurement (POVM) on qubit $B$ \cite{PhysRevLett.113.020402}. Just as the Bloch sphere provides a simple geometric presentation of a single qubit \cite{book1}, the quantum steering ellipsoid gives a faithful and clear presentation for a two-qubit state. Although the phenomenon of quantum steering was noticed at the early days of quantum theory \cite{steer35}, it is quite recently that the construction of QSE for a given two qubit state was thoroughly studied \cite{PhysRevLett.113.020402,steer_njp}. The QSE presentation makes the features of two-qubit states visualized \cite{PhysRevLett.113.020402,PhysRevA.90.024302}, such as the nested tetrahedron condition for separable states, and the radial line segment condition for zero-discord states.

The properties of qubit channels have been the research focus since the beginning of quantum information theory. A local quantum channel acting on one qubit of a two-qubit state can affect the state in various ways. Generally, the local quantum operation decreases the correlation between the two qubits, causing effects such as entanglement sudden death \cite{Yu04}. However, proper local channels can create and increase quantum correlations beyond entanglement, e. g., quantum discord \cite{PhysRevLett.107.170502,PhysRevA.85.032102,PhysRevA.85.010102,PhysRevA.85.022108,PhysRevA.87.032340,PhysRevA.88.012315}. The condition on local channels which can create and increase quantum discord has been well studied \cite{PhysRevLett.107.170502,PhysRevA.85.032102}, yet the set of initial states whose quantum discord can be increased locally is still an open question.

In this paper, we make single-qubit channels visualized, by studying the evolution of the quantum steering ellipsoids of a two-qubit state under a local channel acting on qubit $B$. For both the needle-shape QSE states and the Bell diagonal states with higher dimension QSE, a local channel acting on $B$ shrinks the size of both $\mathcal E_A$ and $\mathcal E_B$. This property can help to judge weather a two-qubit state can be obtained from another by local quantum channels. The QSE $\mathcal E_A^{\Lambda_B}$ of the output state $I\otimes\Lambda_B(\rho_{AB})$ is contained in the scope of $\mathcal E_A$, while QSE of $B$ can be driven away from its original position, which may cause the increase of the $B$-side quantum discord. The connection is built between the the potential of a two-qubit state whose quantum discord can be increased locally and the shape of QSEs for the state. We strictly prove that a $B$-side discordant two-qubit state can be created from a classical state by a trace-preserving local channel if and only if it has needle-shape QSEs $\mathcal E_A$ and $\mathcal E_B$, where $\mathcal E_B$ is not radial. Generally, the quantum discord of a state with higher dimension QSE can be increased by local quantum channels when the shape of QSE is like a baguette, namely, one of the axes length is greatly larger than the two others. Also, we find a class of entangled states whose quantum discord can be increased by local amplitude damping channel.

\section{Effect of Local Channels on States of One-Dimension Quantum Steering Ellipsoids}
Let $\sigma_\mu=\{\sigma_x,\sigma_y,\sigma_z\},\ \mu=1,2,3$ denote the Pauli basis and $\sigma_0=I$ be the single-qubit unitary operator. Any single-qubit state $\rho_A$ can be written as $\rho_A=\frac12\sum_{\mu=0}^3a_\mu\sigma_\mu^A$, where $a_\mu=\mathrm{tr}(\rho_A\sigma_\mu^A)$ and $\boldsymbol a\equiv\{a_1,a_2,a_3\}$ is called the Bloch vector of the state $\rho_A$. Similarly, a two-qubit state $\rho_{AB}$ can be expanded in the Pauli basis as $\rho_{AB}=\frac14\sum_{\mu\nu}\Theta_{\mu\nu}\sigma_\mu^A\otimes\sigma_\nu^B$, where coefficient matrix $\Theta_{\mu\nu}=\mathrm{tr}(\rho_{AB}\sigma_\mu^A\otimes\sigma_\nu^B)$ can be written in the block form
$\Theta=\left(\begin{array}{cc}
1 & \boldsymbol b^\mathrm T \\
\boldsymbol a & T
\end{array}\right)$. Here, $\boldsymbol a$ and $\boldsymbol b$ are the Bloch vectors of the reduced density matrices $\rho_A$ and $\rho_B$ respectively, and the $3\times3$ matrix $T$ represents the correlations.

According to \cite{PhysRevLett.113.020402}, the QSE $\mathcal E_A$ of qubit $A$ can be constructed as follows. Suppose the qubit $B$ is projected to a pure state $\rho_{\boldsymbol x}$, whose Bloch vector is $\boldsymbol x$ with $x=1$. The state of $A$ is steered to $\rho_A^S=\mathrm{tr}_B[\rho_{AB}(I\otimes\rho_{\boldsymbol x})]/\mathrm{tr}[\rho_{AB}(I\otimes\rho_{\boldsymbol x})]$, whose Bloch vector is $\boldsymbol a^S=\frac{\boldsymbol a+T\boldsymbol x}{1+\boldsymbol b\cdot\boldsymbol x}$. For a state with $\boldsymbol b=\boldsymbol 0$, we have $\boldsymbol a^S=\boldsymbol a+T\boldsymbol x$. Therefore, the QSE of qubit $A$ is simply the unit sphere of $\boldsymbol x$, shrunk and rotated by $T$ and translated by $\boldsymbol a$, which can be represented as
\begin{equation}
\mathcal E_A=\{\boldsymbol a+T\boldsymbol x|x\leqslant1\}.\label{QSE}
\end{equation}
For a state with $0<b<1$, one can construct a stochastic local operations and classical communication (SLOCC) operator $S_B\equiv(2\rho_B)^{-1/2}$ on qubit $B$, which takes the two-qubit state to $\rho_{AB}\rightarrow\rho'_{AB}=I\otimes S_B\rho_{AB}(I\otimes S_B)^\dagger$. It is easily checked that $\boldsymbol b'=\boldsymbol 0$, and the QSE $\mathcal E'_A$ of $A$ for state $\rho'_{AB}$ can be obtained following the above method. Since the SLOCC on qubit $B$ does not affect the QSE of qubit $A$, we have $\mathcal E_A=\mathcal E'_A$. Now we are left with the case $b=1$, which means that $\rho_B$ is a pure state and $\rho_{AB}$ must be a product state. Therefore, the QSE of $A$ is simply the single point $\boldsymbol a$. To obtain $\mathcal E_B$, one only need to make the substitution $\boldsymbol a\rightarrow\boldsymbol b,\boldsymbol b\rightarrow\boldsymbol a,T\rightarrow T^\mathrm T$. It is worth mentioning that the QSE $\mathcal E_A$ and $\mathcal E_B$ of state $\rho_{AB}$ have the same dimension, which equals to $\mathrm{rank}(T)=\mathrm{rank}(\Theta)-1$.

An example for states with needle-shape QSE is the so-called classical states, which has zero quantum discord. As proved in Ref. \cite{PhysRevLett.113.020402}, the state $\rho_{AB}$ has zero quantum discord on $B$ if and only if $\mathcal E_B$ is a radial line segment. The general form of a quantum-classical state with zero discord on $B$ is \cite{arXiv:0807.4490v1}
\begin{equation}
\rho_{qc}=p_0\rho_0^A\otimes|\phi_0\rangle_B\langle\phi_0|+p_1\rho_1^A\otimes|\phi_1\rangle_B\langle\phi_1|,\label{classical}
\end{equation}
where $\rho_{0,1}^A$ are linearly independent density matrices of $A$, $|\phi_{0,1}\rangle_B$ are orthogonal pure states of $B$, and $p_0+p_1=1$. Local channels on qubit $B$ can create $B$-side quantum discord from the above quantum-classical state. The output discordant state can be written as
\begin{equation}
\rho\equiv\mathbb I_A\otimes\Lambda_B(\rho_{qc})=p_0\rho_0^A\otimes\rho_0^B+p_1\rho_1^A\otimes\rho_1^B.\label{general_discordant}
\end{equation}
Here $\rho_i^B\equiv\Lambda_B(|\phi_i\rangle_B\langle\phi_i|)$ $(i=0,1)$ do not commute with each other \cite{PhysRevA.84.022113,PhysRevA.85.032102} and thus are linearly independent.

Now we build the connection between locally created discordant states and the states with needle-shape QSE by proving the following statement. \emph{A $B$-side discordant two-qubit state can be created from a classical state by a trace-preserving local channel on $B$ if and only if its QSE of qubit $B$ $\mathcal E_B$ is a non-radial line segment.}

\emph{proof}: We first prove the ``only if'' part, which means that any state $\rho$ in form of Eq. (\ref{general_discordant}) must has needle-shape QSE $\mathcal E_B$ which is not radial. By extending $\rho_i^{A(B)}$ in the Pauli basis as $\rho_i^{A}=\frac12\sum_{\mu=0}^3a_\mu^i\sigma_\mu^A$, $\rho_i^{B}=\frac12\sum_{\mu=0}^3b_\mu^i\sigma_\mu^B$, Eq. (\ref{general_discordant}) can be written as $\rho=\frac14\sum_{\mu\nu}\Theta_{\mu\nu}\sigma_\mu^A\otimes\sigma_\nu^B$. From the linearity of density matrix,
\begin{equation}
\Theta=p_0\Theta^0+p_1\Theta^1,\label{Theta}
\end{equation}
where $\Theta^i=\boldsymbol{a}^i\boldsymbol{b}^{i\mathrm T}$ are coefficient matrices for the product states $\rho_i^A\otimes\rho_i^B$ in the Pauli basis, and hence $\mathrm{rank}(\Theta^0)=\mathrm{rank}(\Theta^1)=1$. According to Eq. (\ref{Theta}), $\mathrm{rank}(\Theta)\leqslant\mathrm{rank}(p_0\Theta^0)+\mathrm{rank}(p_1\Theta^1)=2$. On the other hand, since the discordant state $\rho$ must not be a product state,  we have $\mathrm{rank}(\Theta)>1$. Therefore, $\mathrm{rank}(\Theta)=2$ and the dimensions of both $\mathcal E_A$ and $\mathcal E_B$ equal to $\mathrm{rank}(\Theta)-1=1$. The condition that $\rho$ is discordant on $B$ makes the needle shape $\mathcal E_B$ not radial.

Next we prove that when a state has needle-shape QSE, it can be created locally from a classical state. For a state with one-dimension QSE, the corresponding $\Theta$ matrix is rank 2. Then $\Theta$ can be written as the sum of 2 rank-1 matrices $p_0\Theta^0$ and $p_1\Theta^1$, but not fewer. The rank-1 matrices $\Theta^0$ and $\Theta^1$ stand for product states, so all of the states with needle-shape QSE can be written in the form of Eq. (\ref{general_discordant}). By writing $\rho_0^B$ and $\rho_1^B$ as the spectrum decompositions $\rho_i^B=\lambda_i|\psi_i\rangle\langle\psi_i|+(1-\lambda_i)|\psi_i^\perp\rangle\langle\psi_i^\perp|$, we construct the local channel $\Lambda_B(\cdot)=\sum_{j=0}^3E_j(\cdot)E_j^\dagger$, with $E_0=\sqrt{\lambda_0}|\psi_0\rangle\langle\phi_0|$, $E_1=\sqrt{1-\lambda_0}|\psi_0^\perp\rangle\langle\phi_0|$, $E_2=\sqrt{\lambda_1}|\psi_1\rangle\langle\phi_1|$, and $E_3=\sqrt{1-\lambda_1}|\psi_1^\perp\rangle\langle\phi_1|$, such that $\rho_i^B=\Lambda_B(|\phi_i\rangle_B\langle\phi_i|)$ are satisfied. It means that for any state $\rho$ with needle shape QSE, we can always construct a local channel $\Lambda_B$ which can generate $\rho$ from a quantum-classical state $\rho_{qc}$ as $\mathbb I_A\otimes\Lambda_B(\rho_{qc})=\rho$. This completes the proof.

In order to see explicitly the effect of $\Lambda_B$ on the needle-shape quantum steering ellipsoids $\mathcal E_A$ and $\mathcal E_B$, we study a wide class of two-qubit states in form of
\begin{equation}
\rho=\frac12|+\rangle\langle+|\otimes\rho_B^++\frac12|0\rangle\langle0|\otimes\rho_B^0.\label{rho}
\end{equation}
The general form of $\mathcal E_B$ can be obtained as follows. After the SLOCC operator $S_A=(2\rho_A)^{-1/2}=\frac{\sqrt{2-\sqrt2}}{2}\left(\begin{array}{cc}\sqrt2+1 & -1\\ -1 & \sqrt2+3\end{array}\right)$ acting on $A$, the Bloch vector $\boldsymbol a'$ of the reduced state $\rho'_A$ of $\rho'=S_A\otimes I_B\rho(S_A^\dagger\otimes I_B)$ vanishes, while ${\boldsymbol b}'={\boldsymbol b}=({\boldsymbol b}^{+}+{\boldsymbol b}^0)/2$ does not change. Here ${\boldsymbol b}^{+}$ and ${\boldsymbol b}^0$ are Bloch vectors of $\rho_B^+$ and $\rho_B^0$ respectively. The matrix $T'$ for $\rho'$ is calculated as $T^{\prime\mathrm T}=(\begin{array}{ccc}{\boldsymbol b}^-,\boldsymbol 0,-{\boldsymbol b}^-\end{array})/\sqrt2$, where ${\boldsymbol b}^-\equiv({\boldsymbol b}^{+}-{\boldsymbol b}^0)/2$. From Eq. (\ref{QSE}), the QSE of $B$ is
\begin{eqnarray}
\mathcal E_B&=&\{{\boldsymbol b}'+T^{\prime\mathrm T}{\boldsymbol x}|x\leqslant1\}\nonumber\\
&=&\left\{\boldsymbol b+\frac{ x_1- x_3}{\sqrt2}{\boldsymbol b}^-|x\leqslant1\right\}
\end{eqnarray}
Hence the QSE $\mathcal E_B$ is the segment connecting ${\boldsymbol b}^{+}$ and ${\boldsymbol b}^0$. After the action of channel $\Lambda_B$, the state becomes
\begin{equation}
\rho^\Lambda=\frac12|+\rangle\langle+|\otimes\Lambda_B(\rho_B^+)+\frac12|0\rangle\langle0|\otimes\Lambda(\rho_B^0).\label{rho_lambda}
\end{equation}
Let ${\boldsymbol b}^{+}_\Lambda$ and ${\boldsymbol b}^0_\Lambda$ be the Bloch vector of $\Lambda_B(\rho_B^+)$ and $\Lambda_B(\rho_B^0)$ respectively. Then the QSE $\mathcal E_B^\Lambda$ is the segment connecting  ${\boldsymbol b}^{+}_\Lambda$ and ${\boldsymbol b}^0_\Lambda$ for the output state.

The general form of $\mathcal E_A$ for the state $\rho$ can be derived similarly. The SLOCC operator $S_B=(2\rho_B)^{-1/2}=\frac{1}{\sqrt{(1-b^2)(1+\delta^2)}}(I+\boldsymbol\delta\cdot{\boldsymbol\sigma})$ on qubit $B$ takes the two-qubit state to $\rho''=\frac12|+\rangle\langle+|\otimes\rho_B^{+\prime}+\frac12|0\rangle\langle0|\otimes\rho_B^{0\prime}$. Here $\rho_B^{+/0\prime}=\frac{1}{2(1-b^2)}[(1-{\boldsymbol b}\cdot{\boldsymbol b}^{+/0})I+{\boldsymbol b}^{+/0\prime}\cdot{\boldsymbol\sigma}]$ with ${\boldsymbol b}^{+/0\prime}=\sqrt{1-b^2}{\boldsymbol b}^{+/0}-(1+\boldsymbol\delta\cdot{\boldsymbol b}^{+/0}){\boldsymbol b}$ and $\boldsymbol\delta=-(1-\sqrt{1-b^2})\boldsymbol b/b^2$ are not normalized. Notice that ${\boldsymbol b}^{+\prime}+{\boldsymbol b}^{0\prime}=0$ is satisfied to make sure $\boldsymbol b''=0$. The Bloch vector of reduced state $\rho''_A$ is ${\boldsymbol a}''=\left(\frac{1-{\boldsymbol b}\cdot{\boldsymbol b}^+}{2(1- b^2)},0,\frac{1-{\boldsymbol b}\cdot{\boldsymbol b}^0}{2(1- b^2)}\right)$. since $a''_1+ a''_3=1$, ${\boldsymbol a}''$ is located on the segment $C$ which connects the two points $(1,0,0)^{\mathrm T}$ and $(0,0,1)^{\mathrm T}$ . The $3\times3$ matrix $T''=(\begin{array}{ccc}{\boldsymbol b}^{+\prime},\boldsymbol 0,-{\boldsymbol b}^{+\prime}\end{array})/[2(1-b^2)]$ and consequently $T''{\boldsymbol x}=\frac{{\boldsymbol b}^{+\prime}\cdot{\boldsymbol x}}{\sqrt2(1-b^2)}(\frac{1}{\sqrt2},0,-\frac{1}{\sqrt2})^{\mathrm T}$, which has the same orientation of the segment $C$. Therefore, the QSE $\mathcal E_A$ for state $\rho$ as in Eq. (\ref{rho}) is centered at ${\boldsymbol a}''$ and located on the line segment $C$. The length of $\mathcal E_A$ is
\begin{equation}
l(\mathcal E_A)=\frac{\sqrt2b^{+\prime}}{1-b^2}.\label{la}
\end{equation}
Notice that output state $\rho^\Lambda$ as in Eq. (\ref{rho_lambda}) also belongs to the class described by Eq. (\ref{rho}). After the action of $\Lambda_B$, the QSE $\mathcal E_A^\Lambda$ is still located on the segment $C$.

Before further discussions, we analyze the effect of a local trace-preserving single-qubit channel on the Bloch vector for later convenience. The qubit channel $\Lambda_B$ can operate the Bloch vector ${\boldsymbol b}$ in three ways: \\
(i) It rotates the vector ${\boldsymbol b}$, which corresponds to a unitary channel. \\
(ii) It shrinks the vector ${\boldsymbol b}$ as ${\boldsymbol b}\rightarrow{\boldsymbol b^\Lambda}=\sum_{i=1}^3\lambda_i b_i\vec{\boldsymbol e}_i$, where $|\lambda_i|\leqslant1$ is satisfied to make sure the channel does not increase the trace-norm distance between arbitrary two states. \\
(iii) It translates ${\boldsymbol b}$ by a constant vector as $\boldsymbol b\rightarrow \boldsymbol b^{\tilde\Lambda}=\boldsymbol b+\boldsymbol\lambda_0$. Notice that this type of operation on $\boldsymbol b$ is not a proper channel, since the output vector $\boldsymbol b^{\tilde\Lambda}$ may run out of the Bloch sphere. However, it can be used in combination with channel of type (ii) to make a proper quantum channel. For example, the operation ${\boldsymbol b}\rightarrow{\boldsymbol b^\Lambda}=(1-p){\boldsymbol b}+p\boldsymbol\lambda_0$ represents the channel which destroys the state with probability $p$ and prepares the state of $B$ to a constant qubit state with Bloch vector $\boldsymbol\lambda_0$.

Based on the above calculation, we have the following observations.

(a) Local channel on $B$ does not increase the length of QSE $\mathcal E_B$. This property can be derived from the statement that trace-preserving channels does not increase the trace-norm distance of two states. For states in the form of Eq. (\ref{rho}), the QSE $\mathcal E_B$ is simply the segment connecting the Bloch vectors ${\boldsymbol b}^{+}$ and ${\boldsymbol b}^0$. Hence the length of $\mathcal E_B$ is
\begin{equation}
l(\mathcal E_B)=|{\boldsymbol b}^{+}-{\boldsymbol b}^0|=2D_1(\rho_B^+,\rho_B^0),
\end{equation}
where $D_1(\varrho_1,\varrho_2)\equiv\frac12\mathrm{tr}|\varrho_1-\varrho_2|$ with $|\hat O|\equiv\sqrt{\hat O^\dagger\hat O}$ is the trace-norm distance. Similarly, the length of $\mathcal E_B^\Lambda$ for the output state is $l(\mathcal E_B^\Lambda)=2D_1\left(\Lambda_B(\rho_B^+),\Lambda_B(\rho_B^0)\right)$. Since the channel $\Lambda_B$ can not increase the trace-norm distance of the two states $\rho_B^+$ and $\rho_B^0$, we arrive at $l(\mathcal E_B^\Lambda)\leqslant l(\mathcal E_B)$.

(b) Local channel on $B$ can change the center and orientation of Bob's QSE, and thus has the potential of creating the quantum discord on $B$.
When quantum discord is created by $\Lambda_B$, Bob's QSE change from a radial segment $\overline{\boldsymbol b^+\boldsymbol b^0}$ into a nonradial one $\overline{\boldsymbol b^+_\Lambda\boldsymbol b^0_\Lambda}$. For arbitrary input states $\boldsymbol b^+$ and $\boldsymbol b^0$ located on a radial segment, the channel $\Lambda_B$ can not produce a nonradial segment $\overline{\boldsymbol b^+_\Lambda\boldsymbol b^0_\Lambda}$ if and only if it is one of the following two cases, applied after or followed by a type-(i) channel: Firstly, it is a type-(ii) channel; secondly, it is a combination of type-(ii) channel with $\lambda_1=\lambda_2=0$ and type-(iii) channel with $\boldsymbol\lambda_0=\lambda_0\vec{\boldsymbol e}_3$. For the first case, the channel is a unital channel, and for the second, it is a completely decohering channel. This analysis is in accordance with results in Ref. \cite{PhysRevLett.107.170502} and \cite{PhysRevA.85.032102}. By using the quantum steering ellipsoids, we give a geometric picture of the condition on the local channels with positive quantum-correlating power.

(c) Local channel on $B$ does not increase the length of $\mathcal E_A$. Clearly the unitary channel (i) does not affect $\mathcal E_A$. We briefly prove that the operations of types (ii) and (iii) does not increase the length of $\mathcal E_A$ either. For the local channels of type (ii), we assume that ${\boldsymbol b}=b\vec{\boldsymbol e}_3$ and ${\boldsymbol b}^+=b^+_1\vec{\boldsymbol e}_1+b_3^+\vec{\boldsymbol e}_3$, without loss of generality. In this case, the vector ${\boldsymbol b}^{+\prime}=b_1^+\sqrt{1-b^2}\vec{\boldsymbol e}_1+(b_3^+-b)\vec{\boldsymbol e}_3$. After the action of channel $\Lambda_{ii}^B$, the vector becomes ${\boldsymbol b}^{+\prime}_\Lambda=\lambda_1b_1^+\sqrt{1-(\lambda_3b)^2}\vec{\boldsymbol e}_1+\lambda_3(b_3^+-b)\vec{\boldsymbol e}_3$. According to Eq. (\ref{la}), the length of $\mathcal E_A^\Lambda$ for the output state becomes
\begin{eqnarray}
l(\mathcal E_A^\Lambda)&=&\frac{\sqrt2 b_\Lambda^{+\prime}}{1-(\lambda_3b)^2}\nonumber\\
&=&\left[\frac{2(\lambda_1b_1^{+})^2}{1-(\lambda_3b)^2}+\frac{2\lambda_3^2(b_3^+-b)^2}{\left(1-(\lambda_3b)^2\right)^2}\right]^{\frac12}\nonumber\\
&\leqslant&\left[\frac{2b_1^{+2}}{1-b^2}+\frac{2(b_3^+-b)^2}{(1-b^2)^2}\right]^{\frac12}=l(\mathcal E_A).
\end{eqnarray}
Hence Bob's local channel of type (ii) can not increase the length of QSE of $A$.

For operations of type (iii), we observe that the QSE $\mathcal E_A$ must contain the Bloch vector ${\boldsymbol a}=(\frac12,0,\frac12)^{\mathrm T}$ of $\rho_A$, because Bob can steer $A$ to $\rho_A$ simply by tracing out his qubit. Hence ${\boldsymbol a}$ should locate between the two marginal states ${\boldsymbol a}^+$ and ${\boldsymbol a}^0$. After the action of type-(iii) operation, the state as in Eq. (\ref{rho}) becomes $\rho_{iii}=\rho+\rho_A\otimes\frac12(\boldsymbol\lambda_0\cdot{\boldsymbol\sigma})$. For state $\rho$, let $\hat E^+=\frac12(I+\boldsymbol x\cdot\boldsymbol\sigma)$ be the POVM of $B$ which can steer $A$ to the state $\rho_A^+$ whose Bloch vector is ${\boldsymbol a}^+$ with probability $p^+$. Then for the output state $\rho_{iii}$, the POVM $\hat E^+$ on qubit $B$ steers $A$ to the state ${\boldsymbol a}^+_\Lambda=p{\boldsymbol a}^++(1-p){\boldsymbol a}$ with $p=p^+/(p^++\frac{\boldsymbol\lambda_0\cdot\boldsymbol x}{2})$. It means that the operation of type (iii) draws the marginal state ${\boldsymbol a}^+$ closer to the state ${\boldsymbol a}$. For the same reason, the other marginal state ${\boldsymbol a}^0$ is also drawn closer to ${\boldsymbol a}$. Therefore, the operations of type (iii) can not increase the length of $\mathcal E_A$.  Notice that all of the trace-preserving channels on $B$ can be decomposed as the combination of the operations of the three type, we conclude that the local trace-preserving channel $\Lambda_B$ can not make new states that $A$ can be steered to.

This property can be used to judge weather a state can be obtained from another by local operations. For example, we consider the two states $\rho_1=(1-\delta)(\frac12|+\rangle\langle+|\otimes|1\rangle\langle1|+\frac12|0\rangle\langle0|\otimes|0\rangle\langle0|)+\delta \rho_A\otimes \frac{I}{2}$ and $\rho_2=\frac12|+\rangle\langle+|\otimes|\phi_\delta\rangle\langle\phi_\delta|+\frac12|0\rangle\langle0|\otimes|0\rangle\langle0|$. Here $|\phi_\delta\rangle=\delta|0\rangle+\sqrt{1-\delta^2}|1\rangle$, and $0<\delta\ll1$. The length of QSE for qubit $A$ are $l(\mathcal E_A^1)=\sqrt2(1-\delta)$ and $l(\mathcal E_A^2)=\sqrt2$ respectively. Since $l(\mathcal E_A^1)<l(\mathcal E_A^2)$, the state $\rho_1$ can not be transformed to $\rho_2$ by local operations, even through the mutual information contained in $\rho_1$ is much larger than that contained in $\rho_2$.

\section{Effect of local channels on the quantum steering ellipsoids of Bell diagonal states}
We focus on the Bell diagonal states and study the effect of local channels on the quantum steering ellipsoids of different dimensions. Further, we will study the relation between the effect of locally increased quantum discord and the shape and position of QSE.

For a Bell diagonal two-qubit state, the density matrix can be written as
\begin{equation}
\tilde{\rho}=\frac14\left(\sigma_0\otimes \sigma_0+\sum_{i=1}^3c_i\sigma_i\otimes\sigma_i\right).\label{BellD}
\end{equation}
For such a state, the quantum steering ellipsoids $\mathcal E_A$ and $\mathcal E_B$ are totally the same. Both $\mathcal E_A$ and $\mathcal E_B$ are unit spheres shrunk by $c_1,c_2$ and $c_3$ in the $x, y$ and $z$ direction, respectively.

Now let amplitude damping channel $\Lambda^{\mathrm{AD}}(\cdot)=E_0(\cdot)E_0^\dagger+E_1(\cdot)E_1^\dagger$ acting on qubit $B$, where $E_0=\left(\begin{array}{cc}1 & 0\\0 & \sqrt{1-p}\end{array}\right),E_1=\left(\begin{array}{cc}0 & \sqrt p\\0 & 0\end{array}\right)$. Here $p=1-e^{-\gamma t}$ with $t$ representing the evolution time, and $\gamma$ being the coupling strength between the qubit and the reservoir. It is easily checked that $\Lambda^{\mathrm{AD}}(\sigma_0)=\sigma_0+p\sigma_z,\Lambda^{\mathrm{AD}}(\sigma_{1,2})=\sqrt{1-p}\sigma_{1,2}$, and $\Lambda^{\mathrm{AD}}(\sigma_3)=(1-p)\sigma_3$. Therefore, the amplitude damping channel affect the Bloch vector $\boldsymbol b$ as $\boldsymbol b\rightarrow\boldsymbol b^{\mathrm{AD}}=\sqrt{1-p}(b_1\vec{\boldsymbol e}_1+b_2\vec{\boldsymbol e}_2)+[p+(1-p)b_3]\vec{\boldsymbol e}_3$. It shrinks the Bloch vector by $\sqrt{1-p}$ and $1-p$ in the $x,y$ and $z$ direction respectively, and translates the Bloch vector in the $z$ direction by $p$. Hence the amplitude damping channel is a combination of type (ii) and (iii) operations. After the amplitude damping channel acts on $B$, the Bell diagonal state Eq. (\ref{BellD}) becomes
\begin{eqnarray}
\rho&\equiv& I_A\otimes\Lambda_B(\tilde\rho)\nonumber\\
&=&\frac14[\sigma_0\otimes(\sigma_0+p\sigma_3)+\sqrt{1-p}c_1\sigma_1\otimes\sigma_1\nonumber\\
&&\sqrt{1-p}c_2\sigma_2\otimes\sigma_2+(1-p)c_3\sigma_3\otimes\sigma_3].\label{output}
\end{eqnarray}
As the channel $\Lambda_B^\mathrm{AD}$ affects the coefficient of $\sigma_1\otimes\sigma_1$ and $\sigma_2\otimes\sigma_2$ equivalently, we assume without loss of generality $c_1>c_2$ here and after.

For the output state $\rho$ we first calculate $\mathcal E_A^\mathrm{AD}$. In the limit of $p=1$, $\rho=\frac12\sigma_0^A\otimes|0\rangle_B\langle0|$, and $\mathcal E_A$ is a trivial point at the origin. For $0\leqslant p<1$, let the SLOCC operator $S_B\equiv(2\rho_B)^{-1/2}=\mathrm{diag}(1/\sqrt{1+p},1/\sqrt{1-p})$ acting on qubit $B$, such that $\boldsymbol b'=\boldsymbol0$ for the state $\rho'\equiv I\otimes S_B\rho I\otimes S_B^\dagger$. The corresponding $\boldsymbol a'$ and $T'$ for the state $\rho'$ read $\boldsymbol a'=\left(0,0,-\frac{pc_3}{1+p}\right)^{\mathrm T}$ and $T'=\mathrm{diag}\left(\frac{c_1}{\sqrt{1-p}},\frac{c_2}{\sqrt{1-p}},\frac{c_3}{1-p}\right)$. Hence, we have
\begin{equation}
\mathcal E_A^\mathrm{AD}=\left\{\left(\begin{array}{c}0\\0\\-\frac{pc_3}{1+p}\end{array}\right)+\left(\begin{array}{c}\frac{c_1x_1}{\sqrt{1+p}}\\ \frac{c_2x_2}{\sqrt{1+p}}\\\frac{c_3x_3}{1+p}\end{array}\right)\bigg|x\leqslant1\right\}.
\end{equation}
As $p$ increases from 0, the QSE shrinks with the point $(0,0,-c_3)^\mathrm{T}$ fixed and origin included. It is worth mentioning that, as $p$ gets close but not reaches 1, the correlation between $A$ and $B$ is almost zero, but the QSE $\mathcal E_A^\mathrm{AD}$ still has significate volume. Arbitrary small but positive $1-p$ can ensure $\mathcal E_A^\mathrm{AD}$ with nonzero volume.

For the output state $\rho$ as in Eq. (\ref{output}), the Bloch vector for $\rho_A$ and $\rho_B$ are $\boldsymbol a=\boldsymbol0$ and $\boldsymbol b=(0,0,p)^\mathrm T$ respectively, and the correlation matrix $T=\mathrm{diag}(\sqrt{1-p}c_1,\sqrt{1-p}c_2,(1-p)c_3)$. We can directly write the QSE $\mathcal E_B^\mathrm{AD}$ as
\begin{equation}
\mathcal E_B^\mathrm{AD}=\left\{\left(\begin{array}{c}0\\0\\p\end{array}\right)+\left(\begin{array}{c}\sqrt{1-p}c_1x_1\\ \sqrt{1-p}c_2x_2\\ (1-p)c_3x_3\end{array}\right)\bigg|x\leqslant1\right\}.
\end{equation}
The effect of $\Lambda_B^\mathrm{AD}$ on $\mathcal E_B$ is to translate it by $p$ in the $z$ direction and meanwhile shrink the ellipsoid on three directions. Notice that when $p>\frac{c_3}{1+c_3}$, the ellipsoid does not contain the origin point any more.

On the problem how a local channel affects the QSEs of the Bell diagonal states, we stress the following two points.

(a) Amplitude damping channel acting on qubit $B$ reduces the volume of both $\mathcal E_A$ and $\mathcal E_B$. The QSE $\mathcal E_A^\mathrm{AD}$ reduces to a finite size as $p$ approaches 1 but suddenly become zero when $p=1$. For $c_3=\pm1$ and $c_1,c_2\neq0$, the QSE $\mathcal E_A^\mathrm{AD}$ with $p<1$ does not fit in a tetrahedron that itself fits inside the Bloch sphere. Therefore, the entanglement does not vanish at finite time. This gives an interesting picture for the effect of entanglement sudden death (ESD). At the point where ESD happens, the QSE reduces \emph{smoothly} and comes into a tetrahedron inside the Bloch sphere. However, for the Markovian evolution where the entanglement decreases gradually, there can be a \emph{sudden} change of the QSE volume at $p=1$.

Meanwhile, $\mathcal E_B^\mathrm{AD}$ reduces gradually to zero as $p$ reaches 1. As $\mathcal E_A^\mathrm{AD}$ always contain zero, the quantum steering of $A$ by $B$ is complete during the process of amplitude damping of $B$. For initial states with $c_3=0$ and at least one of $c_1$ and $c_2$ nonzero, the QSE $\mathcal E_B$ is a needle or plate in the $x-y$ plane. When $p>0$, the needle or plate depart from $x-y$ plane, and the affine span of quantum steering ellipsoid of $B$. Therefore, the quantum steering of $B$ become incomplete as long as $p>0$.

(b) Increase of quantum discord by local amplitude damping channel on $B$ can occur for all of the three cases when the initial QSE is a needle, a plate and a ball. For the last two cases, the local increase of discord occurs when $|c_1|\gg |c_2|,|c_3|$, which means that the shape of the plate or the ball is like a baguette perpendicular to the $z$ axis. The quantum discord is calculated following the method in Ref. \cite{PhysRevA.81.042105}.

\begin{figure}
\scalebox{0.23}[0.23]{\includegraphics{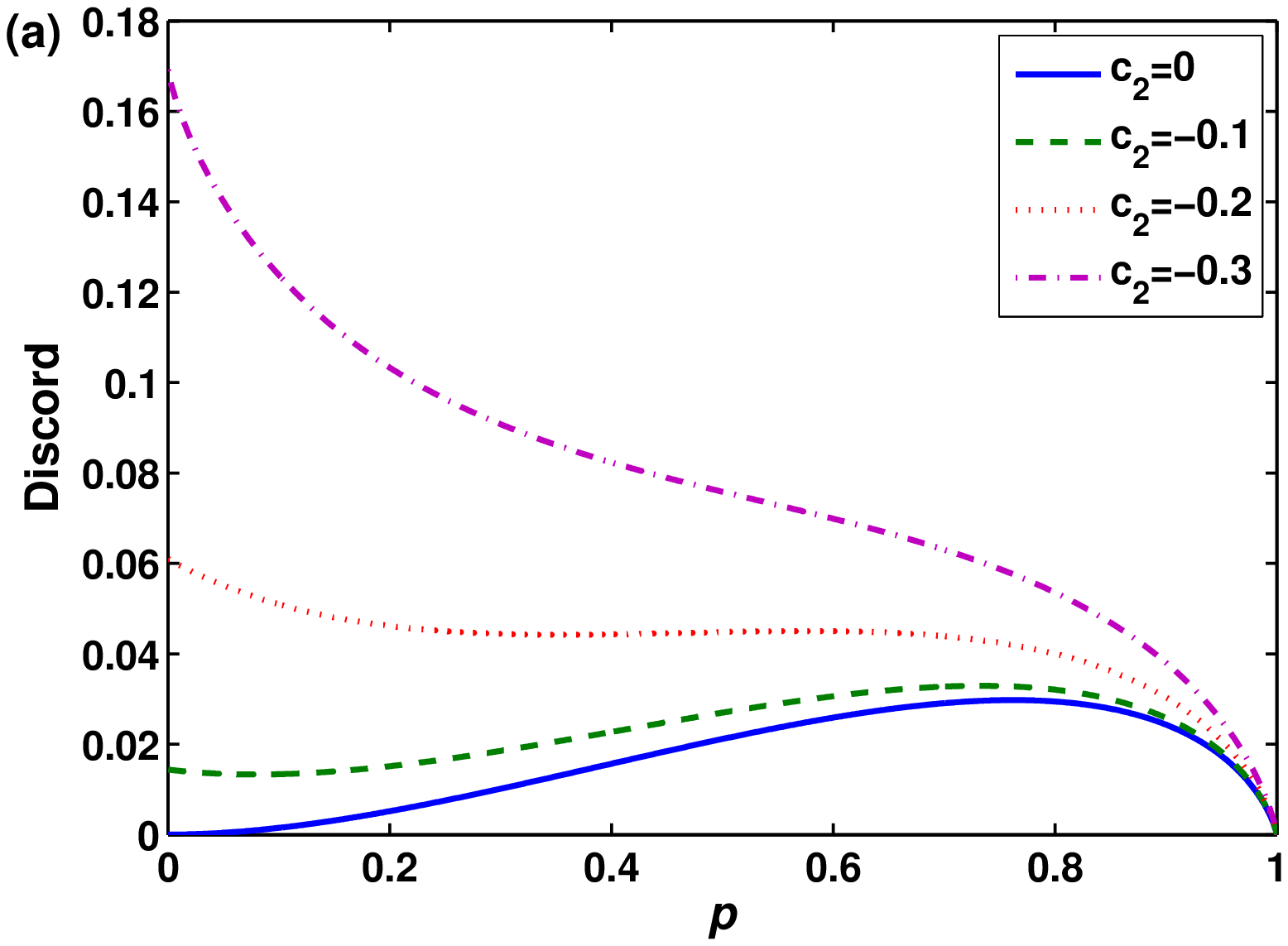}}
\scalebox{0.23}[0.23]{\includegraphics{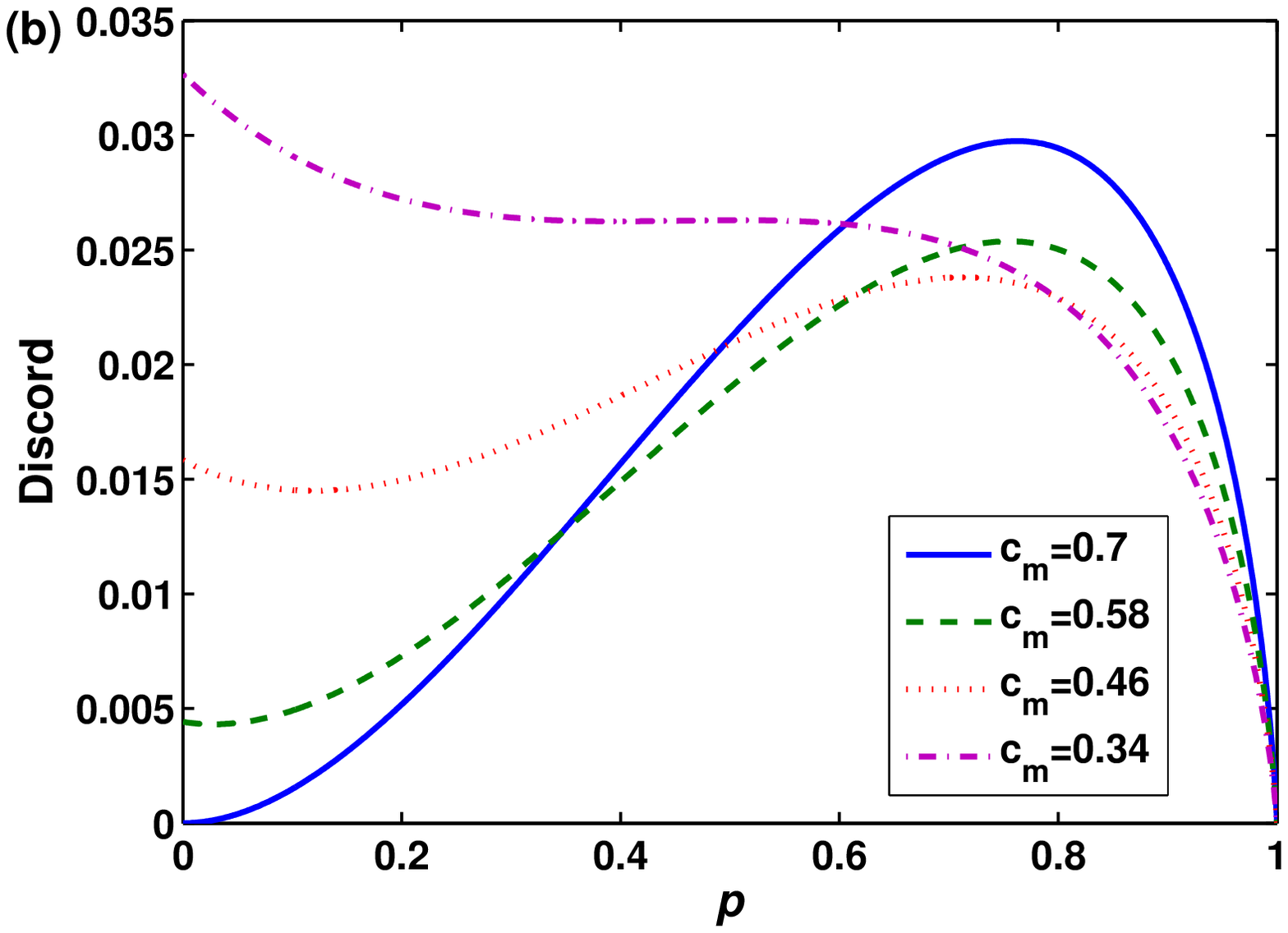}}
\caption{\label{fig1} (Color online) Quantum discord contained in state $\rho$ as in Eq. (\ref{output}) with $c_3=0$ versus the channel parameter $p$. For (a), we fix $c_1=0.7$ and choose $c_2=0,-0.1,-0.2,-0.3$ for different curves. For (b), we fix $c_p\equiv(c_1+c_2)/2=0.7$ and choose $c_m\equiv(c_1-c_2)/2=0.7,0.58,0.46,0.34$ for different curves.}
\end{figure}

In Fig. \ref{fig1} we plot the effect of local amplitude damping channel $\Lambda_B^\mathrm{AD}$ on the quantum discord in the output state $\rho$ as in Eq. (\ref{output}) with $c_3=0$. When $c_2=0$, the QSE $\mathcal E_B^\mathrm{AD}$ is a nonradial needle for $0<p<1$. As shown in the blue solid curve of Fig. \ref{fig1}(a), the quantum discord is created as soon as $p>0$ and does not vanish before $p$ reaches 1. This confirms the statement in Sec. II. When $c_2\neq0$, the quantum steering ellipsoids $\mathcal E_A^\mathrm{AD}$ and $\mathcal E_B^\mathrm{AD}$ are both plates. For $0<p<1$ the affined span of the plate $\mathcal E_B^\mathrm{AD}$ does not contain the origin, and thus the quantum steering of $A$ by $B$ is incomplete. In Fig. \ref{fig1}(a), we fix the value of $c_1$ and increase the value of $c_2$ from zero. As $|c_2|$ increases, the discord in the initial state with $p=0$ gets larger. However, the effect of locally increased quantum discord gets weaker. For the cases $c_2=-0.1$ and $c_2=-0.2$, the quantum discord begins to increase at finite $p$ and the amplitude of the increase is smaller for larger $|c_2|$. For $c_2=-0.3$ the local amplitude damping channel acting on $B$ does not increase the discord for any $p$. In Fig. \ref{fig1}(b), we fix the summation of $c_1$ and $c_2$ but decrease the difference between $|c_1|$ and $|c_2|$. As $c_m$ decreases, the discord of initial state increases, but the maximum discord that $\rho$ can reach is not as large as that for $c_m=c_p$. For $c_m=0.34$, one does not observe the local increase of quantum discord any more. Therefore, we conclude that, for a plate shape QSE, the incompleteness of quantum steering does not always associate with the possibility of locally increased quantum discord. As the plate gets rounder, the effect of locally increased quantum discord tends to vanish.

\begin{figure}
\scalebox{0.23}[0.23]{\includegraphics{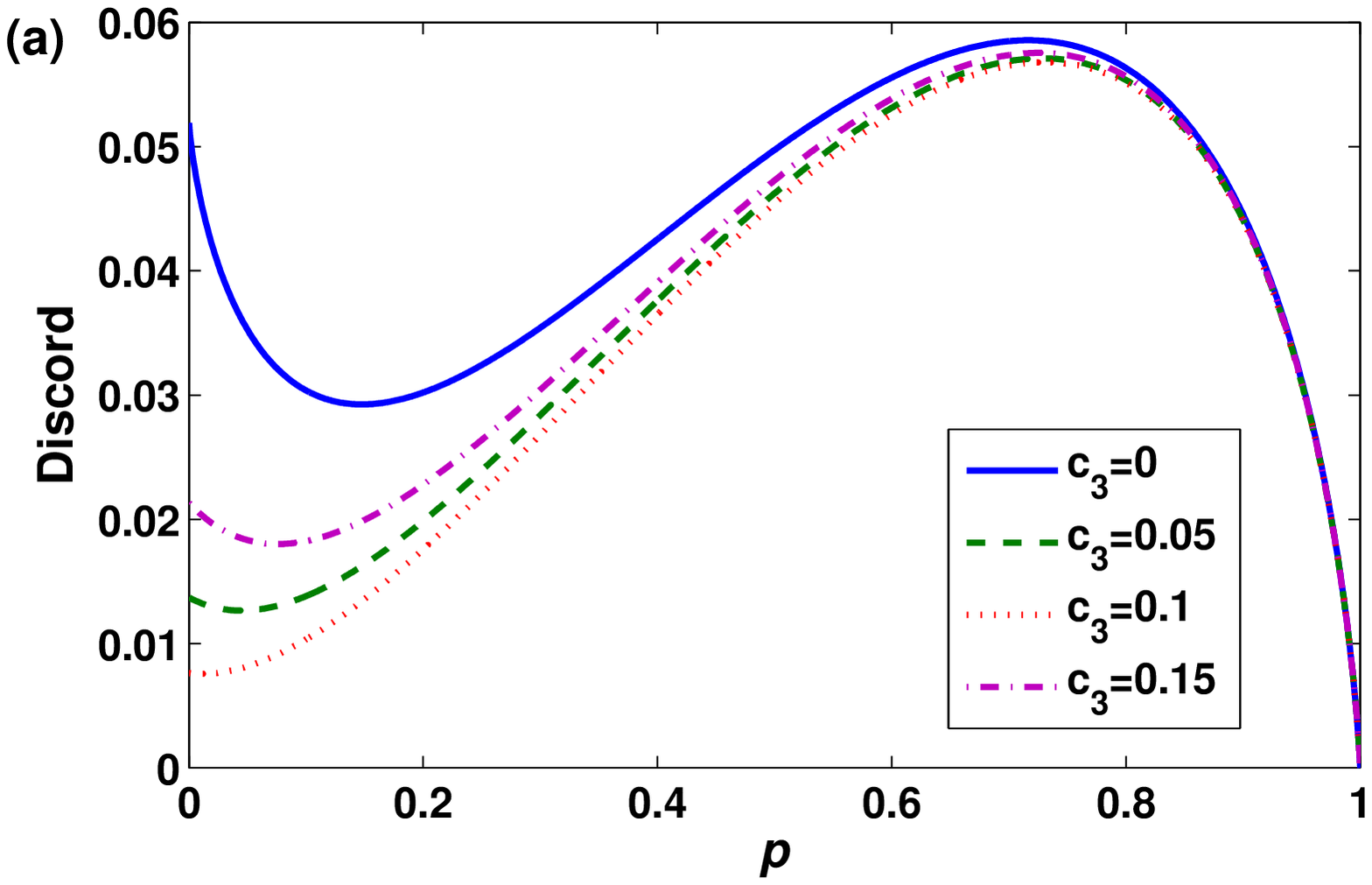}}
\scalebox{0.23}[0.23]{\includegraphics{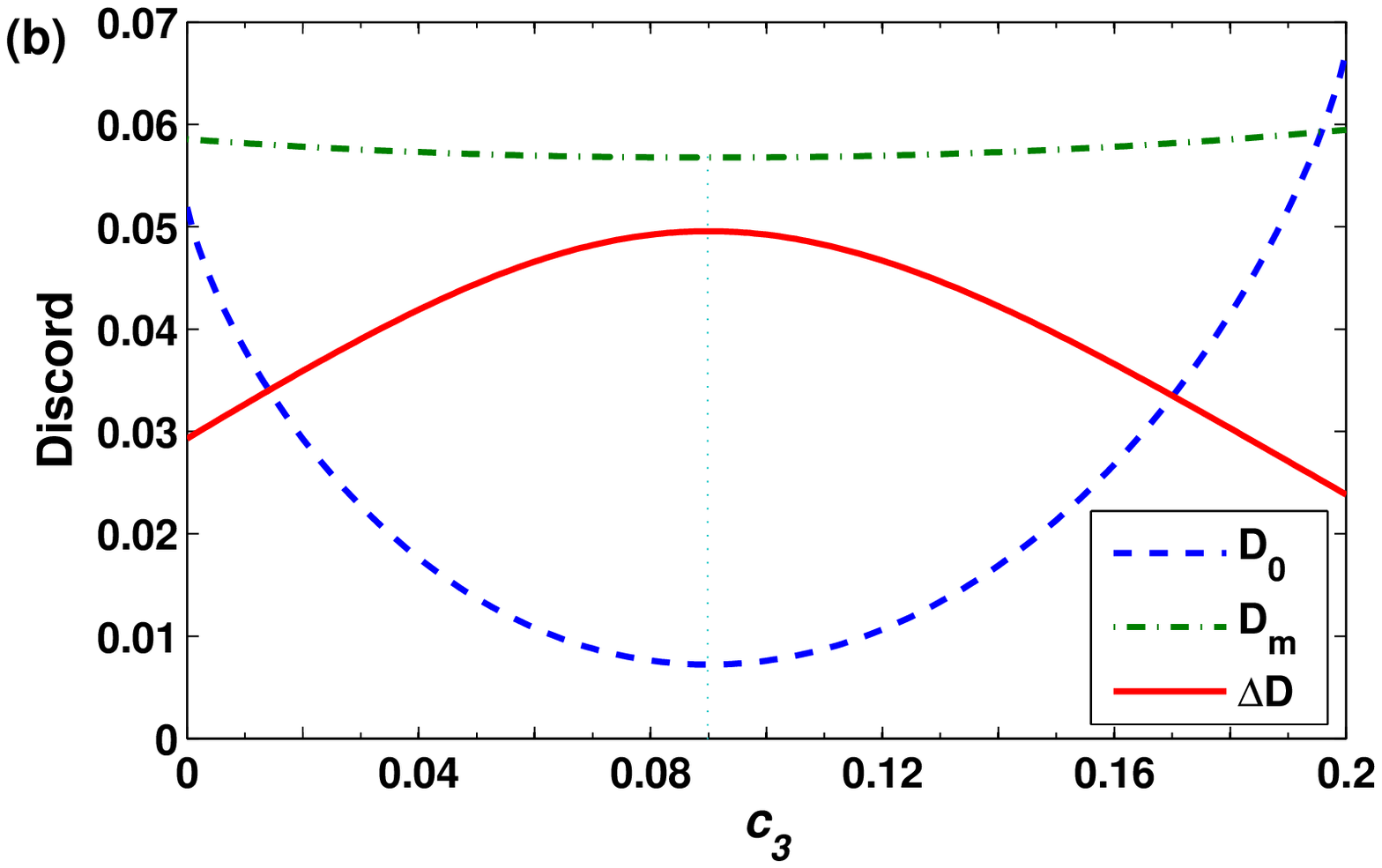}}
\scalebox{0.35}[0.35]{\includegraphics{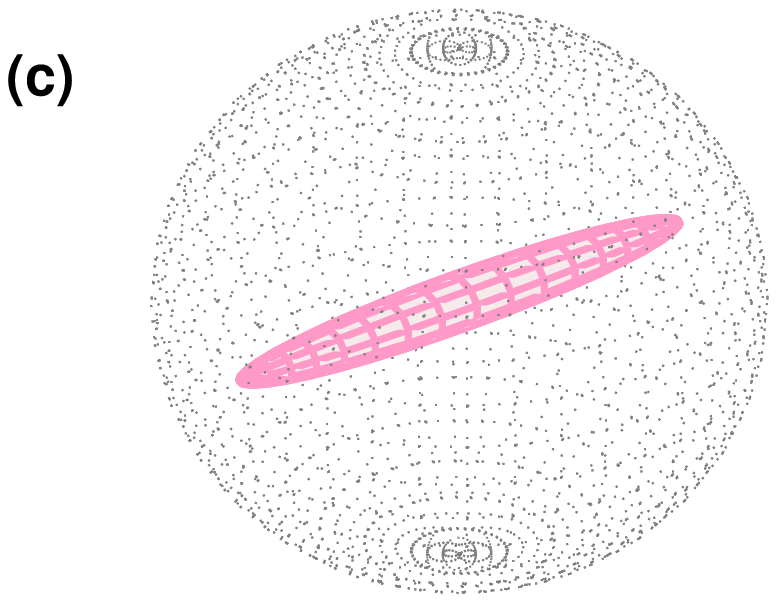}}
\scalebox{0.35}[0.35]{\includegraphics{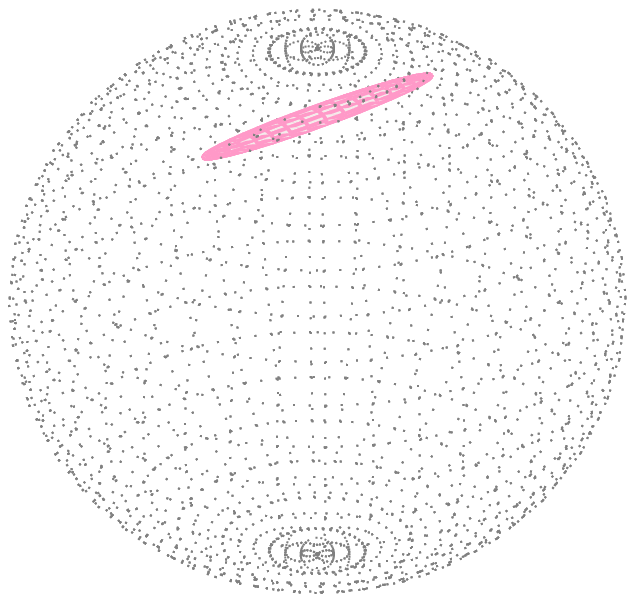}}
\caption{\label{fig2} (Color online) Quantum discord contained in state $\rho$ as in Eq. (\ref{output}) with $c_1=0.9$ and $c_2=-0.1$. (a) Quantum discord versus $p$ for $c_3=0,0.05,0.1,0.15$. (b) The initial quantum discord $D_0$ with $p=0$, the peak value of quantum discord $D_m$ before it vanishes at $p=1$, and the increased quantum discord $\Delta D$, versus $c_3$. (c) The quantum steering ellipsoid of $B$ $\mathcal E_B^\mathrm{AD}$ for $c_3=0.09$ at $p=0.01$(left) and $p=0.73$(right).}
\end{figure}

The effect of locally increased quantum discord for states with three-dimensional QSE is shown in Fig. \ref{fig2}. We fix the value of $c_1=0.9$ and $c_2=-0.1$ and vary $c_3$ to study how the shape of the QSE affects the phenomenon of locally increased quantum discord. From Fig. \ref{fig2}(a), we observe that local amplitude damping channel can increase the quantum discord of states with three-dimension QSE. It means that the effect of locally increased discord does not always associate with incomplete steering. In fact, a nonzero $c_3$ can contribute to the effect of locally increased quantum discord. For $c_3=0.1$, the discord states to increase at smaller $p$, and the increased discord $\Delta D$ is larger, compared with the cases where $c_3=0$, $c_3=0.05$ and $c_3=0.15$. Here $\Delta D$ is defined as the difference between the valley after $p=0$ and the peak value before $p=1$. In Fig. \ref{fig2}(b), we plot the initial discord $D_0$, the peak value $D_m$ and the increased discord $\Delta D$ as functions of $c_3$. The minimization of $D_0$ and $D_m$, as well as the maximization of $\Delta D$, occurs at the same point $c_3=0.09$, which is close to the value of $|c_2|$. For $c_3=0.09$, the quantum discord states to increase at $p=0.01$ and gets to maximum at $p=0.93$. The QSE of $B$ at these two points are shown in Fig. \ref{fig2}(c). It is worth noticing that, at the point $p=0.01$, the concurrence of the two-qubit state as in Eq. (\ref{output}) is $\mathcal C=0.04$. Therefore, the local quantum operation can increase the quantum discord of an entangled state.

\section{Conclusion}
The effect of a local trace-preserving single-qubit channel on the quantum steering ellipsoids of a two-qubit state provides a visualized picture to investigate the phenomena caused by the local channel. The connection is built between a locally induced discordant state and the shape of the QSE of the state. A two qubit state whose QSE of qubit $B$ is a non-radial line segment must be a $B$-side discordant state which can be locally induced from a classical state, and vice versa. For a state with higher-dimensional QSE, its quantum discord has the potential to be increased by local operations usually when the QSE is like a baguette, namely, one of the axes length is greatly larger than the two others. Based on this observation, we find a class of entangled states whose quantum discord can be locally increased. The picture of QSE can help us find the set of two-qubit states whose quantum discord has the potential to be increased locally.

The evolution of the QSE of a two-qubit state under a local channel acting on qubit B has been explicitly studied for both the needle shape QSE and the Bell diagonal states with higher- dimensional QSE. We find that the local channel can not enlarge the size of the QSEs of either qubit $A$ or qubit $B$. This property can help to judge weather a given two-qubit state can be obtained from another by local operations. For the amplitude damping channel acting on qubit $B$, the size of the QSE $\mathcal E_A^\mathrm{AD}$ can have significant size at any finite time $t$, but vanishes for $t\rightarrow\infty$. As long as the ellipsoid before $t$ reaches infinity violates the nested tetrahedron condition, the entanglement does not vanishes at finite time. Therefore, the evolution of QSE under a local channel also provides an interesting picture for the effect of entanglement sudden death.

\begin{acknowledgments}
This work was Supported by the Fundamental Research Funds of Shandong University under grants No. 11170074614037. The National Key Basic Research Program of China under grants NO. 2015CB921003.
\end{acknowledgments}

%\newpage %Just because of unusual number of tables stacked at end

\end{document}